\RequirePackage{fix-cm}
\documentclass[smallextended]{svjour3}       
\smartqed  

\usepackage{hyperref}
\usepackage{graphicx}
\usepackage{float}
\begin{document}
	
	\title{Spectroscopic Study of $\Omega_{cc}$, $\Omega_{cb}$ and $\Omega_{bb}$ Baryons}
	
	
	\author{Amee Kakadiya  \and
		Ajay Kumar Rai
	}
	
	
	\institute{Department of Physics, Sardar Vallabhbhai National Institute of Technology, Surat-395007, Gujarat, India. \at
		\and
		Amee Kakadiya\at
		\email{ameekakadiya@gmail.com}
	}

	\date{Received: date / Accepted: date}

	\maketitle
	
	\begin{abstract}
		In this paper, we presented the mass spectra of the doubly heavy $\Omega$ baryons containing one light (strange) quark and two heavy (charm and bottom) quarks. Our predicted masses can be consider to determine the $J^p$ value for the resonances detected by experimental facilities in future. The masses of ground excited states (1S-6S, 1P-3P, 1D-2D, 1F-2F) are calculated for all possible $J^p$ values, using the Hypercentral Constituent Quark Model (hCQM), by employing screened potential as confining potential with color-Coloumb potential. Regge trajectories are also plotted in $(J, M^2)$ plane for natural and unnatural parities. Doubly heavy Omega states are not declared yet by any experimental facility. We compared our results to the predictions gained from other theoretical approaches, and we found that our predictions are quite close to those of them. Other properties such as magnetic moment (for spin state $\frac{1}{2}$ and $\frac{3}{2}$) and radiative decay width are calculated using the enuerated masses. 
	\end{abstract}
	
	\section{Introduction}
	\label{intro}
	Doubly heavy $\Omega$ baryons provide an opportunity to inspect the behaviour of strange quark in presence of two heavy quarks and also the interaction between two heavy quarks. A serch for singly and doubly heavy baryons was held by well-known experimental colaborations like; SELEX, BABAR, Belle, FOCUS \cite{SELEX,BABAR,Belle2013,Belle2014,FOCUS}. $\Xi_{cc}$ baryon is the only member from doubly heavy family, which is detected experimentally (resonance mass is 3621 MeV) \cite{PDG}. Recently, during $pp$ collission, a bottom-charm baryon with a single light quark is detected in mass range of 6700-7300 $MeV/c^2$ at LHCb experiment, via $\Lambda_c^+ \pi^-$ and $\Xi_c^+ \pi^-$ decay, and they are further decaying into $p K^- \pi^+$ \cite{RAaijOmegaBC}. The resonance may be either $\Xi_{cb}^0$ or $\Omega_{cb}^0$. There is no strong evidence which confirms the isospin of the particle. But, as $\Xi_{cc}$ and $\Xi_{cb}$/$\Omega_{cb}$ are determined experimentally from doubly heavy family, there are major possibilities of other doubly heavy baryons to be found in near future. Our goal of this research is to predict the masses of ground and excited states of doubly heavy $\Omega$ baryons, $\Omega_{cc} (scc)$, $\Omega_{cb} (scb)$ and $\Omega_{bb} (sbb)$; having 0 isospin and -1 strangeness. Furthermore, we study its proprties such as, magnetic moment and radiative decay. The masses are enumerated using Hypercentral Constituent Quark Model (hCQM) \cite{Universe,IJMPA,FBS}, incorporating the screening potential as confining potential with color-Coulomb potential, for quantum numbers: $l=0, 1, 2, 3$, spin-parity: $\frac{1}{2}^+$, $\frac{3}{2}^+$ (natural parity).
	
	Various research groups have been invastigated the doubly heavy baryons with different theoretical approaches such as,  Light-cone sum rules (LCSR) \cite{Aliev2020,AlievPRD}, Bathe-Salpeter equations \cite{Li2020}, Constituent quark model \cite{ShahEPJC2017,ShahEPJC2016,Salehi}, Hamiltonian model \cite{Yoshida}, the variational approach \cite{Roberts}, the three-body Faddeev method \cite{Valcarce}, relativistic approach \cite{Ebert}, Salpeter model \cite{Giannuzzi}, Regge phenomenology \cite{Juhicharm,Juhibottom},  Quark model \cite{Karliner2018}, Heavy baryon chiral perturbation theory \cite{Tong2022},  Lattice QCD \cite{Mathur,MathurCharmBottom}, NRQCD \cite{Mohanta} etc.
	
	This paper is as follows: The theoretical model used to predict the masses of doubly heavy $\Omega$ baryons is described in section \ref{sec:2} after the introduction. The masses of $J^p$ value $\frac{1}{2}$ and $\frac{3}{2}$ and their hyperfine partners are listed in section \ref{sec:3} (see table \ref{tab1}, \ref{tab2}, \ref{tab3}), Regge trajectories are also plotted in ($J, M^2$) plane for natural and unnatural parities. The properties like; ground state magnetic moment, transition magnetic moment and radiative decay width are discussed in section \ref{sec:4}. And conclusion is written in the last section of the article.

	\section{Theoretical Framework}
	\label{sec:2}
	Hypercentral Constituent Quark Model (hCQM) describes the interaction inside the baryonic system containing three quarks. The model is deeply explianed in our previous work \cite{Universe,FBS,IJMPA,ICNFP,triply,ICC2019,DAE2019}, hence brief explaination is given here. The Jacobi coordinates are employed to illustrate the interction between three quarks, which are \cite{Giannini,Giannini2015},
	
	\begin{equation}
		\vec{\rho}=\frac{\vec{r_1}-\vec{r_2}}{\sqrt2} \hspace{0.5cm} and \hspace{0.5cm} 
		\vec{\lambda}=\frac{\vec{r_1}+\vec{r_2}-2\vec{r_3}} {\sqrt{6}}.
		\label{1}
	\end{equation} 
	
	\noindent The hypercentral coordinates (hyperradias $x$ and hyperangle $\xi$) in terms of Jacobi coordinates are \cite{Bijkar2000,Bijkar1994},
	
	\begin{equation}
		x=\sqrt{\rho^2+\lambda^2} 
		\hspace{0.5cm} and \hspace{0.5cm}
		\xi=arctan\left(\frac{\rho}{\lambda}\right)
		\label{2}
	\end{equation}
	
	\noindent The Hamiltonian for the three quark baryonic system is expressed as,
	
	\begin{equation} 
		H=\frac{P^2}{2m} + V(x)
		\label{3}
	\end{equation}
	
	\noindent Where, $P$ is conjugate momentum, $m$ is the reduced mass of the system and  $V(x)$ is non-relativistic interaction potential.
	
	\noindent The expression of kinetic energy operator for the three quark system is \cite{Giannini2015},
	
	\begin{equation} 
		\frac{P^2}{2m}=-\frac{\hbar^2}{2m}(\Delta_{\rho}+\Delta_{\lambda})=-\frac{\hbar^2}{2m}\left(\frac{\partial^2}{\partial x^2}+\frac{5}{x}\frac{\partial}{\partial x}+\frac{L^2(\Omega)}{x^2}\right)
		\label{4}
	\end{equation}
	
	\noindent Here, $L^2(\Omega)$ is the Grand angular operator.

	The predicted mass for particular baryonic state are obtained by solving the Schr\"odinger equation of it, after substituting the $\frac{P^2}{2m}$ term in equation of  Hamiltonian.
	
   The non-relativistic interaction potential inside the baryonic system  $V(x)$ depends on hyperradius x and falls in two terms i.e. spin dependent ($V_{SD}$) and spin independent ($V_{SI}$) potential term \cite{Voloshin2008,IJMPA,FBS}.
	
	\begin{equation}
		V(x)=V_{SD}(x) + V_{SI}(x)
	\end{equation} 
	
	\noindent The spin dependent part of potential $V_{SD}(x)$ is \cite{Thakkar2017},
	\begin{equation}
		V_{SD}(x) = V_{SS}(x)(\vec{S_{\rho}} \cdot \vec{S_{\lambda}}) + V_{\gamma S}(x)(\vec{\gamma} \cdot \vec{S}) +V_T(x) \left[ S^2 - \frac{3 (\vec{S} \cdot \vec{x}) (\vec{S} \cdot \vec{x})}{x^2} \right]
	\end{equation}
	which includes spin-spin, spin-orbit and tensor term respectively. \\ 
	
	And, spin independent part of potential is,
	$V_{SI}(x)=V_{conf}(x) + V_{Col}(x)$ \cite{Gandhi2018,GandhiIJTP2020}. 
	Here, the screened potential is incorporated as confining potential with the color-Coulomb potential, which cab be expressed as,
	
	\begin{equation}
		V_{conf}(x)=a\left(\frac{1-e^{-{\mu} x}}{\mu}\right)  \hspace{0.5cm}  and \hspace{0.5cm} 	V_{Col}(x)= -\frac{2}{3}\frac{\alpha_s}{x}
		\label{5}
	\end{equation}
	
	\noindent where, $a$ is the string tension (different for various baryon system), the constant $\mu$ is the the screening factor, $x$ indicates the inter-quark separation and the parameter $\alpha_s$ corresponds to the strong running coupling constant.
	The values of various parameters used in the calculation are as follows (The values of $\alpha_s$ is given for ground state):

\begin{table}[h]
	\caption{Parameters}
		\begin{center}\label{table0}
		\begin{tabular}{|c|c|c|c|c|c|c|}
			\hline
		$m_s$ & $m_c$ & $m_b$ &	$\mu$ & $\alpha_s (\Omega_{cc})$ & $\alpha_s (\Omega_{cb})$ & $\alpha_s (\Omega_{bb})$\\
		\hline
		0.500 GeV & 1.275 GeV & 4.670 GeV & 0.07 & -0.1230 & -0.0748 & -0.0355 \\
		\hline
		\end{tabular}
	
	\end{center}
\end{table}
 This calculation has been done in the Mathematica notebook \cite{Lucha1999}. The obtained masses are presented in section \ref{sec:3}.
	
	\section{Mass Spectra and Regge Trajectories}
	\label{sec:3}
	In this section, mass spectra of $\Omega_{cc}$, $\Omega_{cb}$ and $\Omega_{bb}$ baryons are presented. And Regge trajectories are also plotted for these baryons using the calculated masses to justify the results. The masses of radial (1S-6S) and orbital (1P-3P, 1D-2D, 1F-2F) states are enumerated using the hCQM (described in section \ref{sec:2}) and listed in Table \ref{tab1}, \ref{tab2} and \ref{tab3}.	
	
	Mass specra of $\Omega_{cc}$ baryon is shown in Table \ref{tab1} and its ground state mass is 3.736 GeV as per our calculation, which is in good accordance with the Refs. \cite{Juhicharm}, \cite{Valcarce} and \cite{Ebert} with the difference of 16 MeV, 39 MeV and 42 MeV respectively and differ by 70-100 MeV with other comparisions given in Table \ref{tab1}. Our 1P state mass is quite near to the Refs. \cite{Valcarce} and \cite{Ebert}, differ by only 2 MeV and 9 MeV respectively. The masses of higher excited states are lower in comapare of other predictions, because screening potential is incorporated as confining potential in our theoretical model.

	For $\Omega_{cb}$ and $\Omega_{bb}$ baryons, our results for 1S state are 7.079 GeV and 10.357 GeV respectively. Ground state mass of $\Omega_{cb}$ baryon is in good agreement with the Refs. \cite{Ebert} and \cite{Tong2022} with difference of only 9 MeV and 1 MeV. The further excited states are also compatible with Refs. \cite{ShahEPJC2016} and \cite{Salehi}. In case of $\Omega_{bb}$ baryon, our 1S state is quite near to the results of Refs. \cite{Ebert} and \cite{Juhibottom} (differ by 2 MeV and 7 MeV respectivel). The result of first orbital state is matching with the results by three-body Fadeev mathod \cite{Valcarce} and relativistic approach \cite{Ebert} with difference of 61 MeV and 41 MeV respectively. The screening effect can be seen in higher excited states. For $\Omega_{bb}$ baryon, the effect of scrrening potential is more than in $\Omega_{cc}$ and $\Omega_{cb}$.

		\begin{table}
		\begin{center}
			\begin{minipage}{\textwidth}
				\caption{Predicted masses of radial and orbital states of $\Omega_{cc}^+$ baryon (in GeV)}\label{tab1}
				\begin{tabular*}{\textwidth}{@{\extracolsep{\fill}}lccccccccccc@{\extracolsep{\fill}}
					}
					\hline\noalign{\smallskip}
					State & Present & \cite{ShahEPJC2016} & \cite{Salehi} & \cite{Yoshida} & \cite{Roberts} & \cite{Valcarce} & \cite{Ebert} & \cite{Giannuzzi} & \cite{Juhicharm} & \cite{MathurCharmBottom}\\
					\noalign{\smallskip}\hline\noalign{\smallskip}
					$1^2S_\frac{1}{2}$ & 3.736 & 3.650 & 3.662 & 3.832 & 3.815 & 3.697 & 3.778 & 3.648 & 3.752 & 3.698\\
					$2^2S_\frac{1}{2}$ & 4.078 & 4.028 & 3.964 & 4.227 & 4.180 & 4.112 & 4.075 & 4.268 \\
					$3^2S_\frac{1}{2}$ & 4.320 & 4.317 & 4.177 & 4.295 &&& 4.321 \\
					$4^2S_\frac{1}{2}$ & 4.514 & 4.570 & 4.452\\
					$5^2S_\frac{1}{2}$ & 4.676 & 4.801 & 4.787\\
					$6^2S_\frac{1}{2}$ & 4.813 \\
					\hline 
					$1^4S_\frac{3}{2}$ & 3.837 & 3.810 & 3.677 & 3.883 & 3.876 & 3.697 & 3.872 & 3.770 & 3.816 \\
					$2^4S_\frac{3}{2}$ & 4.110 & 4.085 & 3.979 & 4.263 & 4.188 & 4.160 & 4.174 & 4.334 \\
					$3^4S_\frac{3}{2}$ & 4.334 & 4.345 & 4.193 & 4.265  \\
					$4^4S_\frac{3}{2}$ & 4.521 & 4.586 & 4.467\\
					$5^4S_\frac{3}{2}$ & 4.680 & 4.811 & 4.802\\
					$6^4S_\frac{3}{2}$ & 4.816 \\ 
					\hline
					\hline
					$1^2P_\frac{1}{2}$ & 4.011 & 3.964 & 3.965 & 4.086 & 4.046 & 4.009 & 4.002 \\
					$1^2P_\frac{3}{2}$ & 4.004 & 3.948 & 3.966 & 4.086 & 4.052 && 4.102 && 3.975\\
					$1^4P_\frac{1}{2}$ & 4.014 & 3.972 & 3.987\\
					$1^4P_\frac{3}{2}$ & 4.007 & 3.981 & 3.978\\
					$1^4P_\frac{5}{2}$ & 3.998 & 3.935 & 3.926 & 4.220 & 4.152 &&&& 4.094\\
					\hline 
					$2^2P_\frac{1}{2}$ & 4.253 & 4.241 & 4.194 & 4.199 & 4.135 & 4.101 & 4.251  \\
					$2^2P_\frac{3}{2}$ & 4.248 & 4.228 & 4.155 & 4.201 & 4.140 && 4.345\\
					$2^4P_\frac{1}{2}$ & 4.256 & 4.248 & 4.215\\
					$2^4P_\frac{3}{2}$ & 4.251 & 4.234 & 4.176\\
					$2^4P_\frac{5}{2}$ & 4.244 & 4.216 & 4.156\\
					\hline 
					$3^2P_\frac{1}{2}$ & 4.453 & 4.492 & 4.453\\
					$3^2P_\frac{3}{2}$ & 4.450 & 4.479 & 4.444\\
					$3^4P_\frac{1}{2}$ & 4.454 & 4.498 & 4.475\\
					$3^4P_\frac{3}{2}$ & 4.451 & 4.486 & 4.466\\
					$3^4P_\frac{5}{2}$ & 4.447 & 4.469 & 4.414\\
					\hline
					\hline
					$1^2D_\frac{3}{2}$ & 4.091 & 4.133 & 4.156\\
					$1^2D_\frac{5}{2}$ & 4.082 & 4.113 & 4.116 & 4.264 & 4.202 &&&& 4.186\\
					$1^4D_\frac{1}{2}$ & 4.101 & 4.156 & 4.215\\
					$1^4D_\frac{3}{2}$ & 4.094 & 4.141 & 4.193\\
					$1^4D_\frac{5}{2}$ & 4.086 & 4.121 & 4.155\\
					$1^4D_\frac{7}{2}$ & 4.075 & 4.095 & 4.086 &&&&&& 4.352\\
					\hline
					$2^2D_\frac{3}{2}$ & 4.318 & 4.389 & 4.429\\
					$2^2D_\frac{5}{2}$ & 4.312 & 4.372 & 4.391& 4.299 & 4.232\\		
					$2^4D_\frac{1}{2}$ & 4.324 & 4.407 & 4.490\\
					$2^4D_\frac{3}{2}$ & 4.320 & 4.395 & 4.468\\
					$2^4D_\frac{5}{2}$ & 4.314 & 4.378 & 4.430\\
					$2^4D_\frac{7}{2}$ & 4.307 & 4.358 & 4.360\\
					\hline
	\hline
$1^2F_\frac{5}{2}$ & 4.157 & 4.287 & 4.391\\
$1^2F_\frac{7}{2}$ & 4.147 & 4.264 & 4.353 &&&&&& 4.387\\
$1^4F_\frac{3}{2}$ & 4.167 & 4.313 & 4.467\\
$1^4F_\frac{5}{2}$ & 4.160 & 4.294 & 4.429\\
$1^4F_\frac{7}{2}$ & 4.150 & 4.271 & 4.360\\ 
$1^4F_\frac{9}{2}$ & 4.139 & 4.244 & 4.292 &&&&&& 4.599\\
\hline
$2^2F_\frac{5}{2}$ & 4.372 & 4.530 & 4.726\\
$2^2F_\frac{7}{2}$ & 4.365 & 4.509 & 4.688\\
$2^4F_\frac{3}{2}$ & 4.380 & 4.552 & 4.802\\
$2^4F_\frac{5}{2}$ & 4.374 & 4.536 & 4.764\\
$2^4F_\frac{7}{2}$ & 4.367 & 4.515 & 4.695\\
$2^4F_\frac{9}{2}$ & 4.359 & 4.490 & 4.627\\
					\hline\noalign{\smallskip}
				\end{tabular*}
			\end{minipage}
		\end{center}
	\end{table}
	
%
%

	\begin{table}[H]
		\begin{center}
			\begin{minipage}{\textwidth}
				\caption{Predicted masses of radial and orbital states of $\Omega_{cb}^0$ baryon (in GeV)}\label{tab2}
				\begin{tabular*}{\textwidth}{@{\extracolsep{\fill}}lcccccccccccc@{\extracolsep{\fill}}
					}
					\hline\noalign{\smallskip}
					State & Present  & \cite{ShahEPJC2016} & \cite{Salehi}& \cite{Roberts} & \cite{Ebert} & \cite{Giannuzzi} & \cite{Karliner2018} & \cite{Tong2022} & \cite{Mathur} & \cite{Mohanta}\\
					\noalign{\smallskip}\hline\noalign{\smallskip}
					$1^2S_\frac{1}{2}$ & 7.079 & 7.136 & 7.329 & 7.136 & 7.088 & 6.994 & 7.013 & 7.078 & 6.994 & 6.946\\
					$2^2S_\frac{1}{2}$ & 7.435 & 7.473 & 7.475 &&& 7.559\\
					$3^2S_\frac{1}{2}$ & 7.688 & 7.753 & 7.609 &&& 7.976\\
					$4^2S_\frac{1}{2}$ & 7.895 & 8.004 & 7.782 \\
					$5^2S_\frac{1}{2}$ & 8.073 & 8.236 & 7.993\\
					$6^2S_\frac{1}{2}$ & 8.227 \\
					\hline 
					$1^4S_\frac{3}{2}$ & 7.182 & 7.187 & 7.339 & 7.187 & 7.130 & 7.017 &&& 7.056\\
					$2^4S_\frac{3}{2}$ & 7.469 & 7.490 & 7.485 &&& 7.571\\
					$3^4S_\frac{3}{2}$ & 7.703 & 7.761 & 7.619 &&& 7.985\\
					$4^4S_\frac{3}{2}$ & 7.903 & 8.009 & 7.792 \\
					$5^4S_\frac{3}{2}$ & 8.077 & 8.239 & 8.002\\
					$6^4S_\frac{3}{2}$ & 8.230 \\ 
					\hline
					\hline
					$1^2P_\frac{1}{2}$ & 7.369 & 7.375 & 7.476\\
					$1^2P_\frac{3}{2}$ & 7.364 & 7.363 & 7.470\\
					$1^4P_\frac{1}{2}$ & 7.371 & 7.381 & 7.490\\
					$1^4P_\frac{3}{2}$ & 7.366 & 7.369 & 7.486\\
					$1^4P_\frac{5}{2}$ & 7.360 & 7.353 & 7.451\\
					\hline
					$2^2P_\frac{1}{2}$ & 7.620 & 7.657 & 7.619\\
					$2^2P_\frac{3}{2}$ & 7.616 & 7.647 & 7.595\\
					$2^4P_\frac{1}{2}$ & 7.621 & 7.662 & 7.633\\
					$2^4P_\frac{3}{2}$ & 7.618 & 7.652 & 7.610\\
					$2^4P_\frac{5}{2}$ & 7.614 & 7.639 & 7.594\\
					\hline
					$3^2P_\frac{1}{2}$ & 7.832 & 7.912 & 7.782\\
					$3^2P_\frac{3}{2}$ & 7.830 & 7.903 & 7.777\\
					$3^4P_\frac{1}{2}$ & 7.833 & 7.916 & 7.796\\
					$3^4P_\frac{3}{2}$ & 7.831 & 7.908 & 7.793\\
					$3^4P_\frac{5}{2}$ & 7.828 & 7.896 & 7.758\\
					\hline
					\hline
					$1^2D_\frac{3}{2}$ & 7.458 & 7.545 & 7.597\\
					$1^2D_\frac{5}{2}$ & 7.452 & 7.531 & 7.571\\
					$1^4D_\frac{1}{2}$ & 7.464 & 7.562 & 7.634\\
					$1^4D_\frac{3}{2}$ & 7.460 & 7.551 & 7.618\\
					$1^4D_\frac{5}{2}$ & 7.454 & 7.536 & 7.595\\
					$1^4D_\frac{7}{2}$ & 7.447 & 7.518 & 7.552\\
					\hline
					$2^2D_\frac{3}{2}$ & 7.695 & 7.807 & 7.768\\
					$2^2D_\frac{5}{2}$ & 7.691 & 7.795 & 7.744\\
					$2^4D_\frac{1}{2}$ & 7.700 & 7.821 & 7.806\\
					$2^4D_\frac{3}{2}$ & 7.697 & 7.812 & 7.792\\
					$2^4D_\frac{5}{2}$ & 7.692 & 7.799 & 7.767\\
					$2^4D_\frac{7}{2}$ & 7.687 & 7.784 & 7.724\\
					\hline
					\hline
					$1^2F_\frac{5}{2}$ & 7.533 & 7.702 & 7.744\\
					$1^2F_\frac{7}{2}$ & 7.526 & 7.685 & 7.720\\
					$1^4F_\frac{3}{2}$ & 7.541 & 7.721 & 7.792\\
					$1^4F_\frac{5}{2}$ & 7.535 & 7.708 & 7.768\\
					$1^4F_\frac{7}{2}$ & 7.528 & 7.690 & 7.723\\
					$1^4F_\frac{9}{2}$ & 7.519 & 7.670 & 7.681\\
					\hline
					$2^2F_\frac{5}{2}$ & 7.758 & 7.949 & 7.956\\
					$2^2F_\frac{7}{2}$ & 7.753 & 7.938 & 7.931\\
					$2^4F_\frac{3}{2}$ & 7.764 & 7.965 & 8.001\\
					$2^4F_\frac{5}{2}$ & 7.760 & 7.953 & 7.978\\
					$2^4F_\frac{7}{2}$ & 7.755 & 7.934 & 7.935\\
					$2^4F_\frac{9}{2}$ & 7.749 & 7.921 & 7.892\\
					\hline\noalign{\smallskip}
					
				\end{tabular*}
			\end{minipage}
		\end{center}
	\end{table}
	
%
%
%
%
%
	
	\begin{table}[H]
		
		\begin{center}
			\begin{minipage}{\textwidth}
				\caption{Predicted masses of radial and orbital states of $\Omega_{bb}^-$ baryon (in GeV)}\label{tab3}
				\begin{tabular*}{\textwidth}{@{\extracolsep{\fill}}lcccccccccccccc@{\extracolsep{\fill}}
					}
					\hline\noalign{\smallskip}
					State & Present  & \cite{ShahEPJC2016} & \cite{Salehi} & \cite{Yoshida} & \cite{Roberts} & \cite{Valcarce} & \cite{Ebert} & \cite{Giannuzzi} & \cite{Juhibottom} & & \\
				\noalign{\smallskip} \hline\noalign{\smallskip}
					$1^2S_\frac{1}{2}$ & 10.357 & 10.446 & 10.870 & 10.447 & 10.454 & 10.293 & 10.359 & 10.271 & 10.350 & & \\
					$2^2S_\frac{1}{2}$ & 10.646 & 10.730 & 10.969 & 10.707 & 10.693 & 10.604 & 10.610 & 10.830 &&& \\
					$3^2S_\frac{1}{2}$ & 10.859 & 10.973 & 11.036 & 10.744 &&& 10.806 & 11.240 &&& \\
					$4^2S_\frac{1}{2}$ & 11.039 & 11.191 & 11.123 & 10.994 &&&&&& \\
					$5^2S_\frac{1}{2}$ & 11.196 & 11.393 & 11.230 &&&&&&& \\
					$6^2S_\frac{1}{2}$ & 11.336 &&&&&&&&& \\
					\hline 
					$1^4S_\frac{3}{2}$ & 10.406 & 10.467 & 10.866 & 10.467 & 10.486 & 10.321 & 10.389 & 10.289 & 10.449 & & \\
					$2^4S_\frac{3}{2}$ & 10.662 & 10.737 & 10.964 & 10.723 & 10.721 & 10.622 & 10.645 & 10.622 &&& \\
					$3^4S_\frac{3}{2}$ & 10.866 & 10.976 & 11.032 & 10.730 &&& 10.843 & 11.247 &&& \\
					$4^4S_\frac{3}{2}$ & 11.043 & 11.193 & 11.119 & 11.031 &&&&&& \\
					$5^4S_\frac{3}{2}$ & 11.199 & 11.394 & 11.225 &&&&&&& \\
					$6^4S_\frac{3}{2}$ & 11.337 &&&&&&&&& \\
					\hline 
					\hline
					$1^2P_\frac{1}{2}$ & 10.580 & 10.634 & 10.968 & 10.607 & 10.616 & 10.519 & 10.532 &&& \\
					$1^2P_\frac{3}{2}$ & 10.578 & 10.629 & 10.957 & 10.608 & 10.619 & 10.520 & 10.566 && 10.638 && \\
					$1^4P_\frac{1}{2}$ & 10.581 & 10.636 & 10.976 &&&&&&& \\
					$1^4P_\frac{3}{2}$ & 10.579 & 10.631 & 10.963 &&&&&&& \\
					$1^4P_\frac{5}{2}$ & 10.576 & 10.625 & 10.956 &&&&&& 10.729 && \\
					\hline 
					$2^2P_\frac{1}{2}$ & 10.796 & 10.881 & 11.031 & 10.796 & 10.763 & 10.683 & 10.738 &&& \\
					$2^2P_\frac{3}{2}$ & 10.795 & 10.877 & 11.029 & 10.797 & 10.765 & & 10.775 &&& \\
					$2^4P_\frac{1}{2}$ & 10.797 & 10.883 & 11.039 &&&&&&& \\
					$2^4P_\frac{3}{2}$ & 10.796 & 10.879 & 11.035 &&&&&&& \\
					$2^4P_\frac{5}{2}$ & 10.794 & 10.874 & 11.019 &&&&&&& \\
					\hline 
					$3^2P_\frac{1}{2}$ & 10.982 & 11.104 & 11.124 &&&&&&& \\
					$3^2P_\frac{3}{2}$ & 10.981 & 11.101 & 11.111 &&&&&&& \\
					$3^4P_\frac{1}{2}$ & 10.982 & 11.106 & 11.131 &&&&&&& \\
					$3^4P_\frac{3}{2}$ & 10.981 & 11.103 & 11.119 &&&&&&& \\
					$3^4P_\frac{5}{2}$ & 10.980 & 11.098 & 11.112 &&&&&&& \\
					\hline
					\hline
					$1^2D_\frac{3}{2}$ & 10.660 & 10.783 & 11.032 &&&&&&& \\
					$1^2D_\frac{5}{2}$ & 10.657 & 10.777 & 11.017 & 10.729 & 10.720 &&&& 10.918 & & \\
					$1^4D_\frac{1}{2}$ & 10.662 & 10.789 & 11.038 &&&&&&& \\
					$1^4D_\frac{3}{2}$ & 10.661 & 10.785 & 11.033 &&&&&&& \\
					$1^4D_\frac{5}{2}$ & 10.658 & 10.779 & 11.019 &&&&&&& \\
					$1^4D_\frac{7}{2}$ & 10.655 & 10.772 & 11.007 &&&&&& 11.002 & &\\
					\hline
					$2^2D_\frac{3}{2}$ & 10.864 & 11.012 & 11.116 &&&&&&& \\
					$2^2D_\frac{5}{2}$ & 10.863 & 11.008 & 11.104 &&&&&&& \\
					$2^4D_\frac{1}{2}$ & 10.866 & 11.017 & 11.126 &&&&&&& \\
					$2^4D_\frac{3}{2}$ & 10.865 & 11.014 & 11.119 &&&&&&& \\
					$2^4D_\frac{5}{2}$ & 10.863 & 11.009 & 11.106 &&&&&&& \\
					$2^4D_\frac{7}{2}$ & 10.861 & 11.004 & 11.094 &&&&&&& \\
					\hline
					\hline
					$1^2F_\frac{5}{2}$ & 10.727 & 10.920 & 11.105 &&&&&&& \\
					$1^2F_\frac{7}{2}$ & 10.724 & 10.913 & 11.082 &&&&&& 11.191 && \\
					$1^4F_\frac{3}{2}$ & 10.730 & 10.927 & 11.118 &&&&&&& \\
					$1^4F_\frac{5}{2}$ & 10.728 & 10.922 & 11.107 &&&&&&& \\
					$1^4F_\frac{7}{2}$ & 10.725 & 10.915 & 11.094 &&&&&&& \\
					$1^4F_\frac{9}{2}$ & 10.721 & 10.907 & 11.076 &&&&&& 11.268 & & \\
					\hline
					$2^2F_\frac{5}{2}$ & 10.923 & 11.136 & 11.210 &&&&&&& \\
					$2^2F_\frac{7}{2}$ & 10.921 & 11.130 & 11.189 &&&&&&& \\
					$2^4F_\frac{3}{2}$ & 10.925 & 11.142 & 11.226 &&&&&&& \\
					$2^4F_\frac{5}{2}$ & 10.923 & 11.137 & 11.214 &&&&&&& \\
					$2^4F_\frac{7}{2}$ & 10.921 & 11.132 & 11.202 &&&&&&& \\
					$2^4F_\frac{9}{2}$ & 10.919 & 11.125 & 11.179 &&&&&&& \\
					\hline\noalign{\smallskip}
					
				\end{tabular*}
			\end{minipage}
		\end{center}
	\end{table}
%
%
%

\begin{figure}
	\begin{minipage}{.45\textwidth}
		\includegraphics[width=0.9\textwidth]{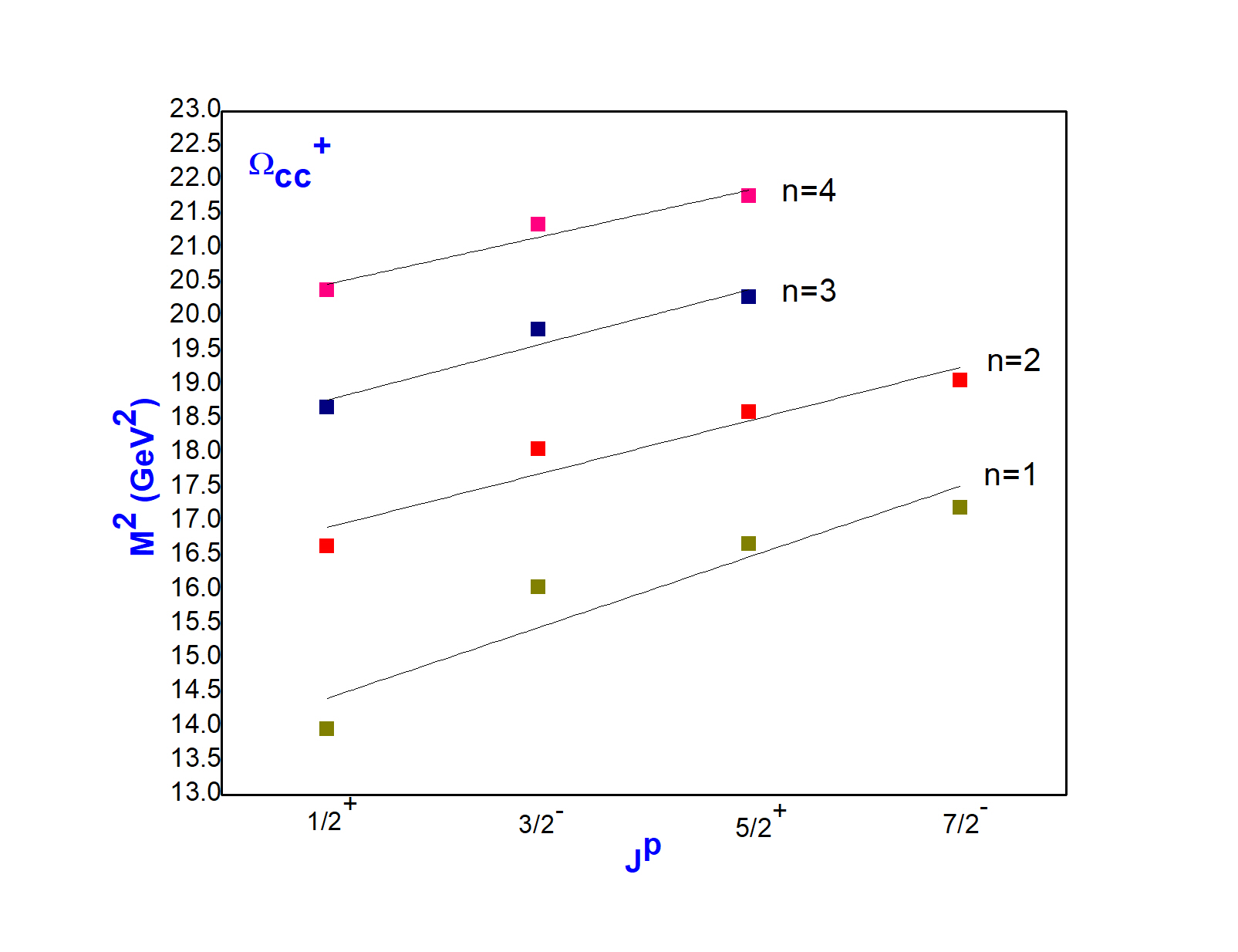}
		\caption{The $M^2 \rightarrow J$ Regge trajectory of $\Omega_{cc}^{+}$ baryon with natural parity}\label{fig2}
	\end{minipage}
	\hfill  
	\begin{minipage}{.45\textwidth}
		\includegraphics[width=0.7\textwidth]{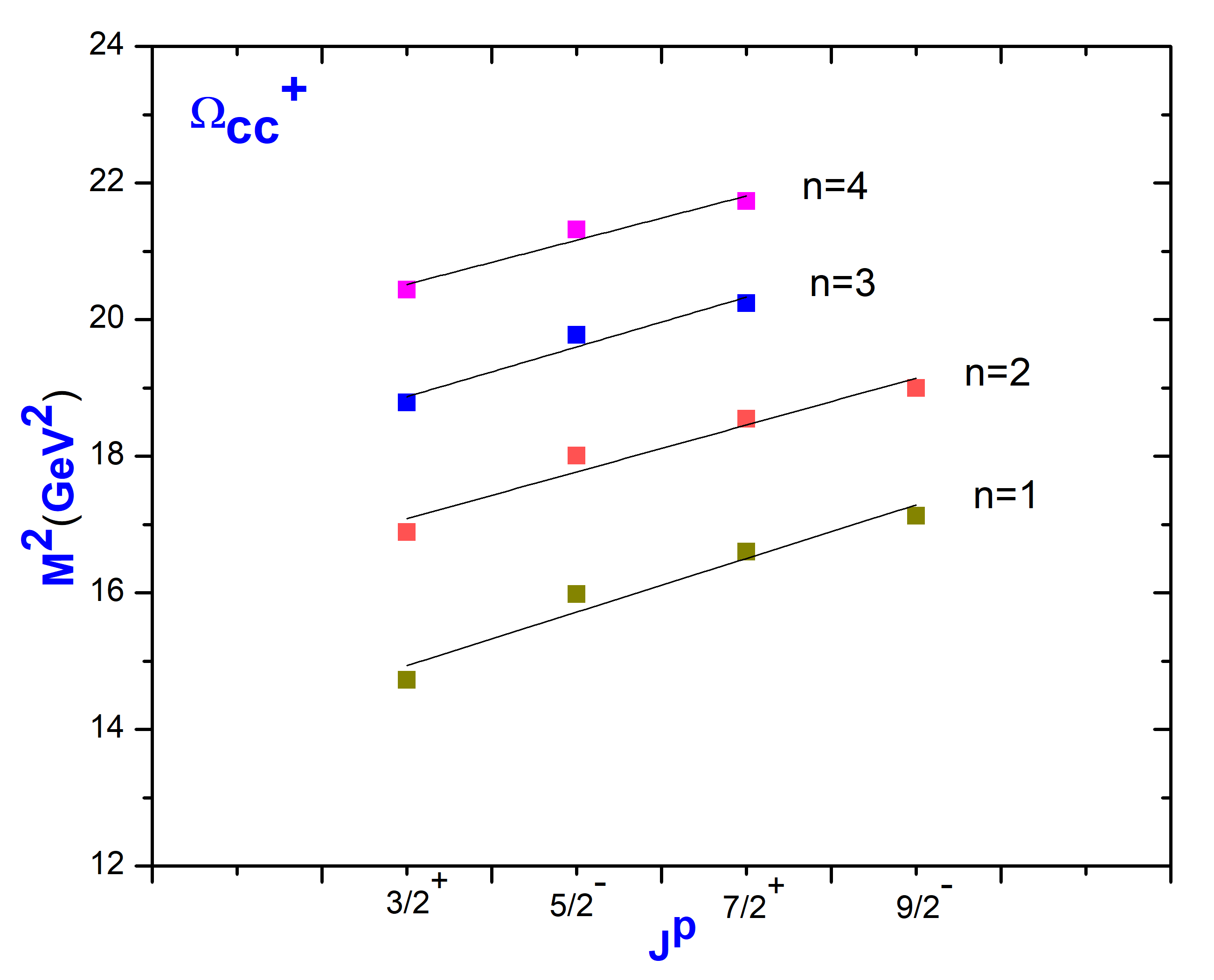}
		\caption{The $M^2 \rightarrow J$ Regge trajectory of $\Omega_{cc}^{+}$ baryon with unnatural parity}\label{fig3}
	\end{minipage}
\end{figure}
\begin{figure}
	\begin{minipage}{.45\textwidth}
		\includegraphics[width=0.9\textwidth]{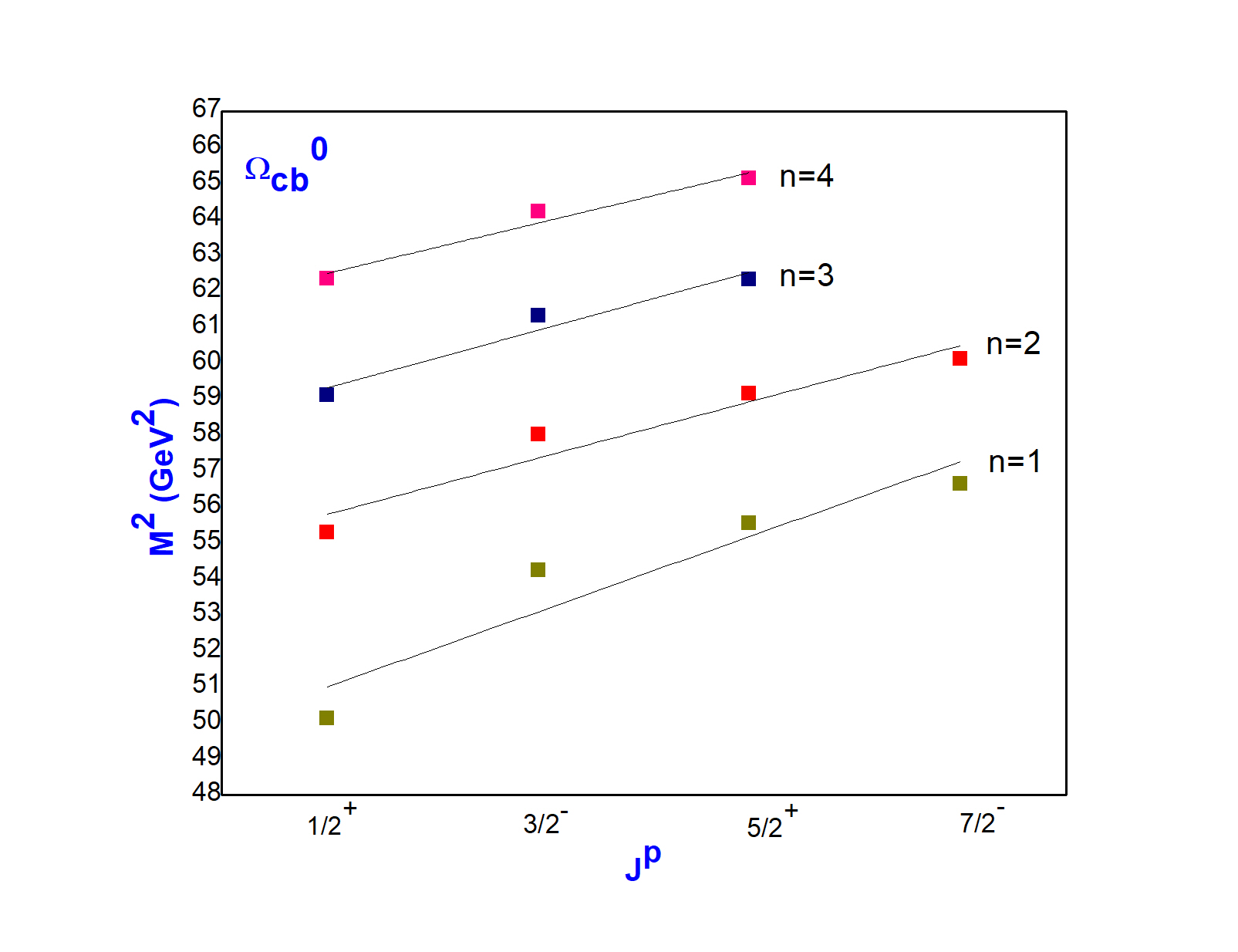}
		\caption{The $M^2 \rightarrow J$ Regge trajectory of $\Omega_{cb}^{0}$ baryon with natural parity}\label{fig4}
	\end{minipage}
	\hfill  
	\begin{minipage}{.45\textwidth}
		\includegraphics[width=0.7\textwidth]{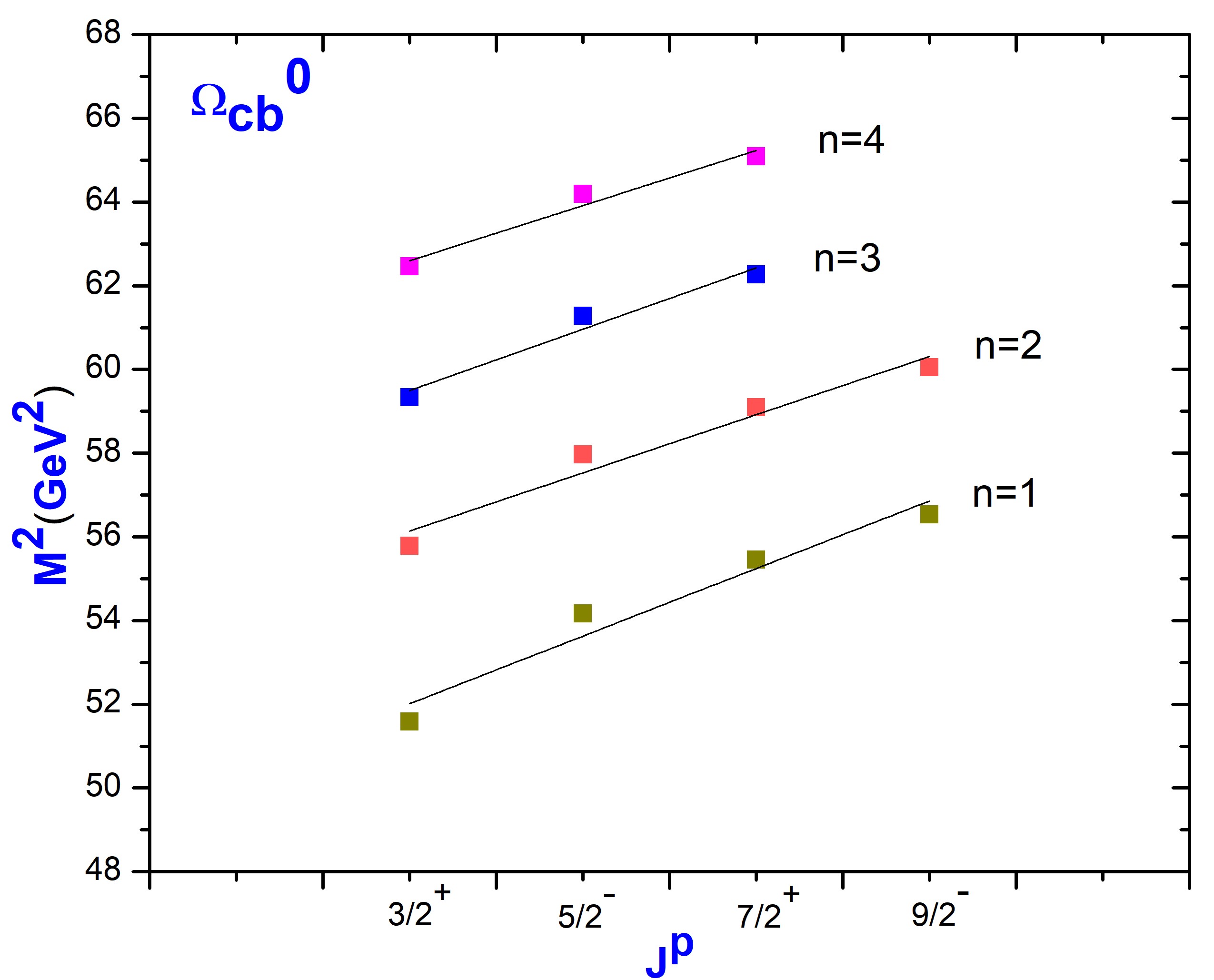}
		\caption{The $M^2 \rightarrow J$ Regge trajectory of $\Omega_{cb}^{0}$ baryon with unnatural parity}\label{fig5}
	\end{minipage}
\end{figure}
\begin{figure}
	\begin{minipage}{.45\textwidth}
		\includegraphics[width=0.9\textwidth]{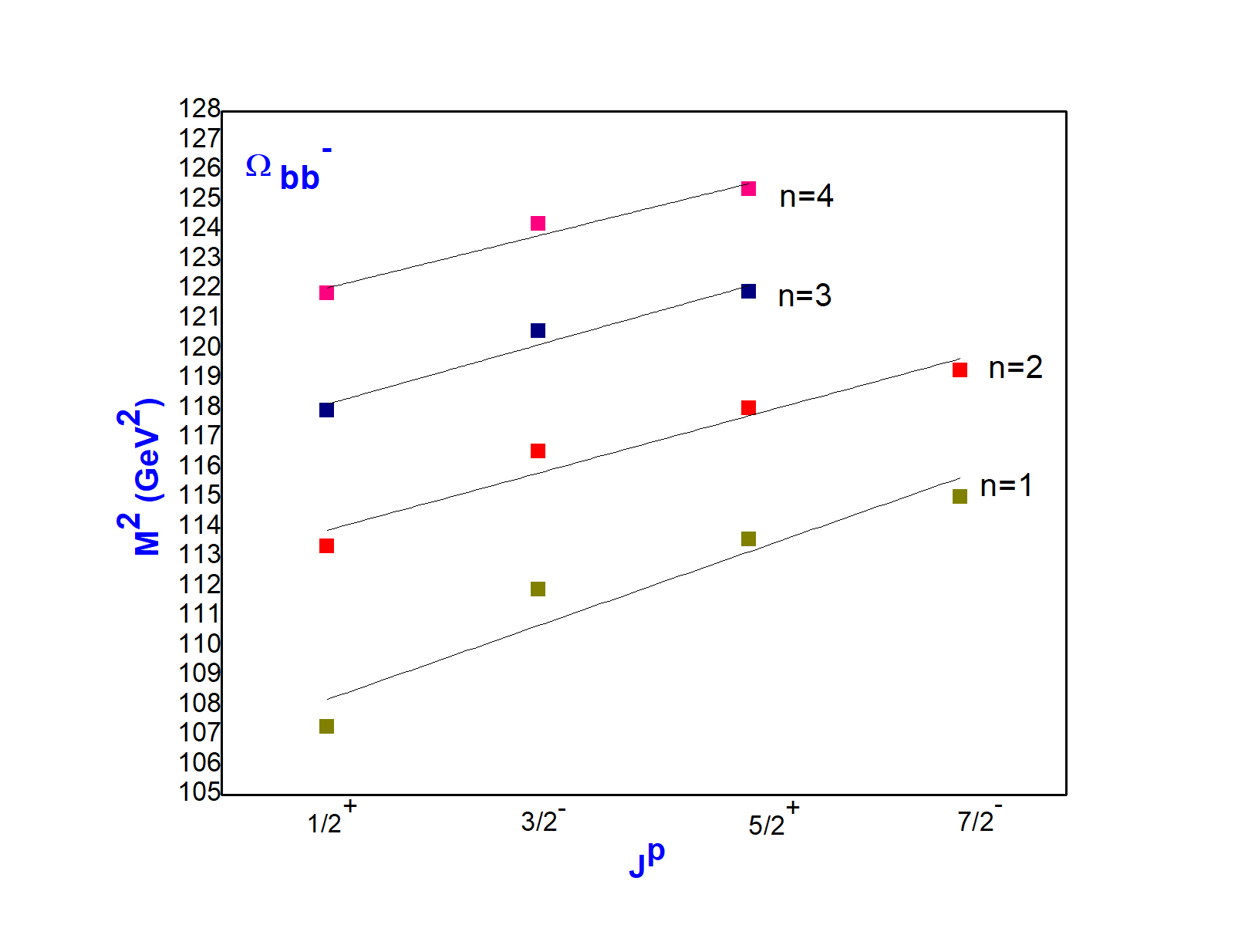}
		\caption{The $M^2 \rightarrow J$ Regge trajectory of $\Omega_{bb}^{-}$ baryon with natural parity}\label{fig6}
	\end{minipage}
	\hfill  
	\begin{minipage}{.45\textwidth}
		\includegraphics[width=0.7\textwidth]{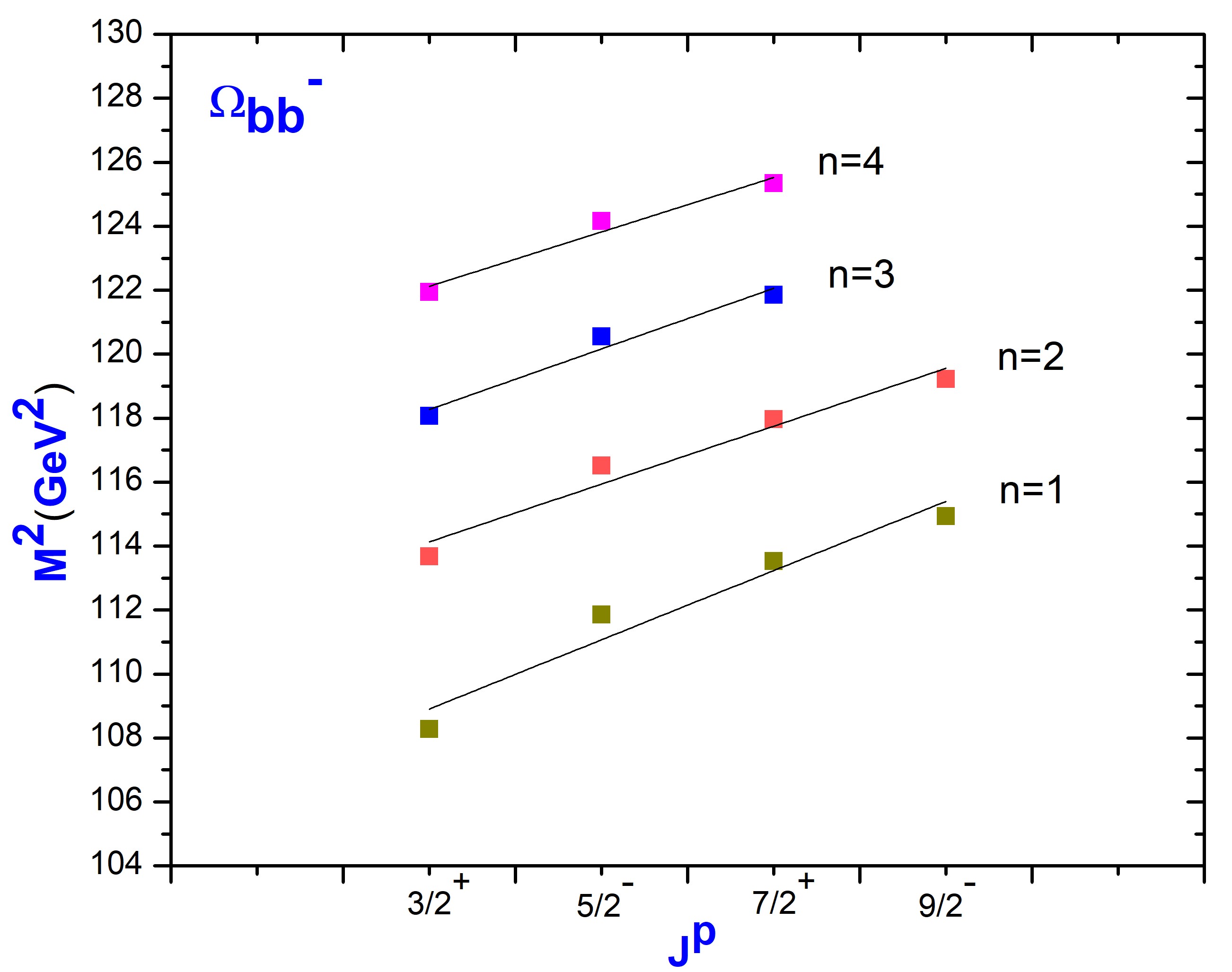}
		\caption{The $M^2 \rightarrow J$ Regge trajectory of $\Omega_{bb}^{-}$ baryon with unnatural parity}\label{fig7}
	\end{minipage}
\end{figure}

	Regge trajectories of $\Omega_{cc}$, $\Omega_{cb}$ and $\Omega_{bb}$ for natural and unnatural parities are shown in Figure \ref{fig2} - \ref{fig7}. They are plotted in $(J,M^2)$ plane, using calculated masses presented in Table \ref{tab1}, \ref{tab2} and \ref{tab3}. Our plots are shown in F The nature of Regge trajectories for baryons is suppose to be linear.  Here, the Regge lines are linear at first sight, but for higher excited states beyond the $J^p$ value $\frac{7}{2}^-$, they are going to intersect each other. This is caused by the confining potential (screening potential) used in our theoretical model, called screening effect.

	\section{Magnetic moment and Radiative decay}
	\label{sec:4}
	\subsection{Magnetic Moment}
	Magnetic moment is an intrinsic property of a baryon, which can be determine by using spin, charge and effective mass of its constituent quarks. The expression of magnetic moment for a baryon can be obtain by operating the spin-flavour wave function to the $z$-component of the magnetic moment operator \cite{GandhiDecay,BPatel},\\
	
	\begin{equation}
		\mu_{B}=\sum_{q} \langle \Phi_{sf} \|\hat{\mu}_{qz} \|\Phi_{sf} \rangle    \quad \quad q=s, c, b
	\end{equation}
	
	Here, $\Phi_{sf}$ represents the spin-flavour wave-function of the baryon and $\hat{\mu}_{qz}$ is the magnetic moment operator. \\
	\noindent The magnetic moment of individual quark is given by,
	\begin{equation}
		\mu_q=\frac{e_q}{2m_q^{eff}}\cdot \sigma_q
	\end{equation}
	
	where, $e_q$ and $\sigma_q$ are charge and spin of the individual constituent quark of the baryonic system respectively and $m_q^{eff}$ is the effective mass of constituent quark which is expressed as given below \cite{GandhiDecay}:
	\begin{equation}
		m_q^{eff}=m_q \left(1+\frac{\langle H \rangle}{\sum_q m_q}\right)
	\end{equation}
	Here, $\langle H \rangle=M-\sum_q m_q$; where, $M$ is predicted mass of the particular baryonic state. And $m_q^{eff}$ is the mass of bounded quark inside the baryon with consideration of the interaction with other two quarks. \\
	The magnetic moment of spin states $\frac{1}{2}$ and $\frac{3}{2}$ for $\Omega_{cc}$, $\Omega_{cb}$ and $\Omega_{bb}$ baryons are calculated using above equations and masses presented in Section \ref{sec:3}. Our results are shown in the Table \ref{tab4} and compared with the other magnetic moment results obtained by different approaches. Our calculated magnetic moment results are differ by 0.02-0.52 $\mu_N$ from Ref. \cite{ShahEPJC2016}, by 0.007-0.4 $\mu_N$ from Ref. \cite{BernotasPRD}, by 0.004-0.1 $\mu_N$ from Ref. \cite{BPatel}, by 0.01-0.1 from Ref. \cite{Albertus} and by 0.01-1 $\mu_N$ from Ref. \cite{DhirPRD2021}.

	\begin{table}[h]
		\begin{center}
			\begin{minipage}{\textwidth}
				\caption{Magnetic moment of ground state of doubly heavy $\Omega$ baryons (in $\mu_N$)}\label{tab4}
				\begin{tabular*}{\textwidth}{@{\extracolsep{\fill}}lcccccccccccc@{\extracolsep{\fill}}}
					\hline\noalign{\smallskip}
					Baryon& $J^p$ & Expression & Present & \cite{ShahEPJC2016} & \cite{BernotasPRD} & \cite{BPatel} & \cite{Albertus} & \cite{DhirPRD2021} \\
					\noalign{\smallskip}\hline\noalign{\smallskip}
					$\Omega_{cc}^+$ & $\frac{1}{2}^+$ & $\frac{4}{3}\mu_c-\frac{1}{3}\mu_s$ & 0.704 & 0.692 & 0.668 & 0.785 & 0.635 & 0.7109 \\
					&&&&&&&& $\pm 0.0017$ \\
					$\Omega_{cb}^0$ & $\frac{1}{2}^+$ & $\frac{2}{3}\mu_c+\frac{2}{3}\mu_b-\frac{1}{3}\mu_s$  & 0.447 & 0.439 & 0.034 & 0.397 & 0.368 & -0.6202 \\
					&&&&&&&& $\pm 0.00001$ \\
					$\Omega_{bb}^-$ & $\frac{1}{2}^+$ & $\frac{4}{3}\mu_b-\frac{1}{3}\mu_s$ & 0.113 & 0.108 & 0.120 & 0.109 & 0.101 & 0.1135  \\
					&&&&&&&& $\pm 0.0008$ \\
					$\Omega_{cc}^{*+}$ & $\frac{3}{2}^+$ & $2\mu_c+ \mu_s$ & 0.283 & 0.285 & 0.332 & 0.121 & 0.139 & 0.1871 & \\
					&&&&&&&& $ \pm 0.0026$ \\
					$\Omega_{cb}^{*0}$ & $\frac{3}{2}^+$ & $\mu_c+ \mu_b+ \mu_s$ & -0.181 & -0.181 & -0.111 & -0.317 &-0.261 & -0.2552  \\
					&&&&&&&& $\pm 0.0016$ \\
					$\Omega_{bb}^{*-}$ & $\frac{3}{2}^+$ & $2\mu_b+ \mu_s$ & -0.718 & -1.239 & -0.730 & -0.711 & -0.662 & -0.6999 \\
					&&&&&&&& $\pm  0.0017$ \\
					\hline\noalign{\smallskip}
				\end{tabular*}
			\end{minipage}
		\end{center}
	\end{table}

	\noindent 
	\subsection{Radiative Decay}
	The $\gamma$ emission while changing the spin state $\frac{3}{2}$ to $\frac{1}{2}$ is known as `Radiative decay'. The radiative decay width is expressed as \cite{GandhiDecay,Shah2016cpc},
	\begin{equation}
		\Gamma=\frac{k^3}{4\pi}\frac{2}{2J+1}\frac{e^2}{2m_p^2}\mu_{B\rightarrow B'}^2
	\end{equation}
	where, $k$ is photon energy, $J$ is total angular momentum of the initial baryonic state, $m_p$ is the mass of proton (in MeV) and $\mu_b$ transition magnetic moment for the particular radiative decay.
	
	The transition magnetic moment for the $\gamma$ emission, can be calculated by the sandwiching spin-flavour wave functions of initial baryon state($B$) and final baryon state($B'$) with $z$ component of magnetic moment operator, which is expressed as below:
	\begin{equation}
		\mu_{B \rightarrow B'}=\langle \Phi_B \|\mu_{B\rightarrow B'} \|\Phi_{B'}\rangle    
	\end{equation}
	The spin-flavour wave function of initial baryon($\Phi_B$) state and final baryon state($\Phi_B'$) can be determined as described in  \cite{Majethiya2009}.
	
	The transition magnetic moment and radiative decay widths for doubly heavy $\Omega$ baryons are shown in Table \ref{tab5} and compared with the results of Ref. \cite{Li} and \cite{DhirPRD2021}.

	\begin{table}[h]
		\begin{center}
			\begin{minipage}{\textwidth}
				\caption{Transition magnetic moment (in $\mu_N$) and radiative decay width (in keV) of doubly heavy $\Omega$ baryons}\label{tab5}
				\begin{tabular*}{\textwidth}{@{\extracolsep{\fill}}lcccccccccc@{\extracolsep{\fill}}}
					\hline\noalign{\smallskip}
					Transition & Expression & \multicolumn{3}{c}{Transition magnetic moment}   &\multicolumn{2}{c}{Radiative decay width}&  \\
					&  & Present &  \cite{Li} & \cite{DhirPRD2021} & Present  & \cite{DhirPRD2021} \\
					\noalign{\smallskip}\hline\noalign{\smallskip}
					$\Omega_{cc}^{*+}\rightarrow \Omega_{cc}^+ \gamma$ & $\frac{\sqrt{2}}{3}(\mu_s-\mu_b)$ & 0.847 &0.96 &0.9111 & 2.947 & 1.973  \\
					&&&& $\pm 0.0013$ && $\pm 0.029$\\
					$\Omega_{cb}^{*0}\rightarrow \Omega_{cb}^0 \gamma$ & $\frac{\sqrt{2}}{3}(\mu_s-\mu_b)$ & 0.713 &0.69 &-0.7896 & 2.255 & 0.579 \\
					&&&& $\pm 0.0010$ && $\pm 0.014 $\\
					$\Omega_{bb}^{*-}\rightarrow \Omega_{bb}^- \gamma$ & $\frac{\sqrt{2}}{3}(\mu_s-\mu_b)$ & 0.499 &0.48 &0.4906 & 0.121 & 0.0426  \\	
					&&&& $\pm 0.0014$ && $\pm 0.0018$\\
					\hline\noalign{\smallskip}
				\end{tabular*}
			\end{minipage}
		\end{center}
	\end{table}

	\section{Conclusion}
	\label{sec:5}
	The mass spectra of doubly heavy omega baryons with strangeness -1 for all possible hyperfine states are determined in this work. To enumereate the masses of radial and orbital states, Hypercentral Constituent Quark Model (hCQM) has been utilized. In model, screening potential has been incorporated as a confining part with the color-Coulomb potential. Mass spectra of $\Omega_{cc}$, $\Omega_{cb}$ and $\Omega_{bb}$ baryons have been enumerated by solving six-dimentional Schr\"odinger equation in Mathematica notebook.   Our results are in good agreement with the results from various theoretical approaches \cite{ShahEPJC2016,Salehi,Yoshida,Roberts,Valcarce,Ebert,Giannuzzi,Juhicharm,Juhibottom,Karliner2018,Tong2022,Mathur,MathurCharmBottom,Mohanta}. For $\Omega_{cc}$ baryon, the recent result obtained by Lattice QCD calculation \cite{MathurCharmBottom} is quite matching with our ground state with difference of 38 MeV. The first orbital state is only differ by 47 MeV from Ref. \cite{ShahEPJC2016}. After 3P state, the screening effect is visible in $\Omega_{cc}$ baryon. Looking into the results of $\Omega_{cb}$ baryon, our ground state is only 34 MeV differ from Refs. \cite{Mathur} and \cite{Roberts}, which are the results obtained using lattice QCD and variational approach respectively. Our 1P states is only 6 MeV less than Ref. \cite{ShahEPJC2016}. From 3P state onwards, masses are going to suppress due to screened potential. The ground state of $\Omega_{bb}$ baryon is only 7 MeV lower than Ref. \cite{Juhibottom}. Here, screening is visible after 2P state. Though doubly heavy baryon states are not experimentally established, we have no experimental proof to justify our results. Our goal is to offer a solid theoretical evidence to the resonances that will be discovered experimentally in the near future. The results calculated in this paper will be helpful to determine $J^p$ value of newly detected experimental resonances.
	
	After the mass spectra calculation, Regge trajectories are plotted in $(J, M^2)$ plane using the enumerated masses for $\Omega_{cc}$, $\Omega_{cb}$ and $\Omega_{bb}$ baryons. The nature of Regge trajectories validate the obtained masses. Regge lines follow the linearity, but  effect Regge lines slightly seems contracting rather than being parallel, due to screening effect.
	
	 The properties like ground state magnetic moment, transition magnetic moment and radiative decay width for $\Omega_{cc}$, $\Omega_{cb}$ and $\Omega_{bb}$ baryons are determined using the calculated mass spectra. The results of ground state magnetic moment are compared to Refs. \cite{ShahEPJC2016,BernotasPRD,BPatel,Albertus,DhirPRD2021} and found quite near to them. Transition magnetic moment and radiative dacay width are also calculated and compared to Ref. \cite{DhirPRD2021} and \cite{Li}. Our results are in accordance with the compared results.


	%
	%


\begin{thebibliography}{}
		%
		%
		
		
		\bibitem{SELEX} S. Koshkarev et al. (SELEX Collaboration), Phys. Lett. B 765, 171 (2017).
		
		\bibitem{BABAR} B. Aubert et al. (BABAR Collaboration), Phys. Rev. D \textbf{74}, 011103 (2006). 
		
		\bibitem{Belle2013}  R. Aaij et al. (Belle Colaboration), J. High Energy Phys. \textbf{12}, 090 (2013).
		
		\bibitem{Belle2014} Y. Kato et al. (Belle Collaboration), Phys. Rev. D \textbf{89}, 052003 (2014).
		
		\bibitem{FOCUS} S. P.  Ratti et al. (FOCUS Collaboration), Nucl. Phys. Proc. Suppl. \textbf{115}, 33 (2003).
		
		\bibitem{PDG} R.L. Workman et al. (Particle Data Group), Prog. Theor. Exp. Phys. 2022, 083C01 (2022).
		
		\bibitem{RAaijOmegaBC} R. Aaij, et al., Chin. Phys. C \textbf{45}, 093002 (2021).
		
		\bibitem{Giannini}M. M. Giannini, E Santopinto, Hypercentral Constituent Quark Model, AIP Conf.Proc. 1488 no.1, \textbf{257-265}(2012).
		
		\bibitem{Giannini2015} M.M. Giannini and E. Santopinto, {\it Chin. J. Phys.} {\bf 53}, 020301 (2015).
		
		\bibitem{Li2009} B. Q. Li and K. T. Chao, {\it Phys. Rev. D} {\bf 79}, 094004 (2009).
		
		\bibitem{Aliev2020} T. M. Aliev and , K. Simsek, Eur. Phys. J. C \textbf{80}, 976 (2020). 
		
		\bibitem{AlievPRD} H. I. Alrebdi, T. M. Aliev and , K. Simsek,  Phys. Rev. D \textbf{102}, 074007 (2020).
		
		\bibitem{Li2020} Q. Li, C. H. Chang, S. X. Qin and G. L. Wang, Chin. Phys. C  \textbf{44}, 013102  (2020).
		
		\bibitem{ShahEPJC2017} Z. Shah and A. K. Rai, Eur. Phys. J. C  \textbf{77}, 129 (2017).
		
		\bibitem{ShahEPJC2016} Z. Shah, K. Thakkar, A. K. Rai, Eur. Phys. J. C  \textbf{76}, 530 (2016).
		
		\bibitem{Salehi} N. Salehi, Acta Phys. Polon. B \textbf{50}, 735-752 (2019).
		
		\bibitem{Yoshida}  T. Yoshida, E. Hiyama, A. Hosaka, M. Oka, K. Sadato, Phys. Rev. D \textbf{92}, 114029 (2015).
		
		\bibitem{Roberts} W. Roberts, M. Pervin, Int. J. Modern Phys. A \textbf{23}, 2817 (2008).
		
		\bibitem{Valcarce} A. Valcarce, H. Garcilazo, J. Vijande, Eur. Phys. J. A \textbf{37}, 217 (2008).
		
		\bibitem{Ebert} D. Ebert, R.N. Faustov, V.O. Galkin, A.P. Martynenko, Phys. Rev. D \textbf{66}, 014008 (2002).
		
		\bibitem{Giannuzzi} F. Giannuzzi, Phys. Rev. D \textbf{79}, 094002 (2002).
		
        \bibitem{Juhicharm} J. Oudichhya, K. Gandhi, and A. K. Rai, Phys. Rev. D \textbf{103}, 114030 (2021).
		
		\bibitem{Juhibottom} J. Oudichhya, K. Gandhi, and A. K. Rai, Phys. Rev. D \textbf{104}, 114027 (2021).
		
		\bibitem{Karliner2018} M. Karliner and J. L. Rosner, Phys. Rev. D \textbf{97}, 094006 (2018).
		
		\bibitem{Tong2022} H. Z. Tong and H. S. Li, arXiv:2110.01380v2 [hep-ph]
		
		\bibitem{Mathur} N. Mathur, M. Padmanath, and S. Mondal, Phys. Rev. Lett. \textbf{121}, 202002 (2018).
		
		\bibitem{MathurCharmBottom} N. Mathur, R. Lewis and R. M. Woloshyn, Phys. Rev. D \textbf{66}, 014502 (2002).
		
		\bibitem{Mohanta} P. Mohanta and S. Basak, Phys. Rev. D \textbf{101}, 094503 (2020).
		
		\bibitem{Universe} Z. Shah, A. Kakadiya,  K. Gandhi, A.K. Rai, \textit{Universe} \textbf{7}, 337 (2021).
		
		\bibitem{FBS} A. Kakadiya, Z. Shah, K. Gandhi and A. K. Rai, Few-Body Systems \textbf{63}, 29 (2022).
		
		\bibitem{IJMPA} A. Kakadiya, Z. Shah and A. K. Rai, Int. Jour. of Mod. Phys. A \textbf{37}, No. 11n12, 2250053 (2022).
		
        \bibitem{ICNFP} A. Kakadiya, C. Menapara, A. K. Rai,  arXiv:2204.13438 [hep-ph], (2022).
        
        \bibitem{triply} A. Kakadiya, Z. Shah and A. K. Rai, Int. Jour. of Mod. Phys. A, https://doi.org/10.1142/S0217751X22502256.
		
		\bibitem{ICC2019} K. Gandhi, A. Kakadiya, Z. Shah, and A. K. Rai,   Proc. 3rd Int. Conf. on Condenced Matter and Applied Physics (AIP Conf. Proc. \textbf{2220} (2020)), 140015.
		
		\bibitem{DAE2019} A. Kakadiya, K. Gandhi, A. K. Rai, Orbitally excitation of $\Lambda_b^0$ baryon, in Proc. 64th DAE BRNS Symposyum on Nuclear Physics, (Lucknow, Uttar Pradesh, 2019), p.~697.
		
		\bibitem{Bijkar2000}  R. Bijkar, F. Iachello, A. Leviatan, Ann. Phys. \textbf{284}, 89 (2000).
		
		\bibitem{Bijkar1994}  R. Bijkar, F. Iachello, A. Laviatan, Ann. Phys. (N. Y.) \textbf{236}, 69 (1994).
		
		\bibitem{Voloshin2008} M.B. Voloshin, Prog. Part. Nucl. Phys. \textbf{61}, 455 (2008).
		
		\bibitem{Thakkar2017}K. Thakkar, Z. Shah, A.K. Rai and P.C. Vinodkumar, Nucl. Phys. A \textbf{965}, 57 (2017).
		
		\bibitem{Gandhi2018} K. Gandhi, Z. Shah and A. K. Rai, Eur. {\it Phys. J. Plus} {\bf 133}, 512 (2018).
		
		\bibitem{GandhiIJTP2020}  K. Gandhi, Z. Shah and  A.K. Rai, Int. J. Theor. Phys. \textbf{59}, 1129–1156 (2020).
		
		\bibitem{Lucha1999}W. Lucha and F. Schoberls, Int. J. Mod. Phys. C \textbf{10}, 607 (1999).
		
		\bibitem{GandhiDecay}
		K. Gandhi, Z. Shah, A. K. Rai, Eur. Phys. J. Plus \textbf{133}, 512 (2018).
		
		\bibitem{BPatel} B. Patel, A.K. Rai, P.C. Vinodkumar, J. Phys. G \textbf{35}, 065001 (2008).
		
		\bibitem{Shah2016cpc}Z. Shah, K. Thakkar, A.K. Rai and P.C. Vinodkumar, Chin. Phys. C \textbf{40}, 123102 (2016).
		%
		%
		
		
		
		
		\bibitem{Majethiya2009}
		A. Majethiya, B. Patel, P. C. Vinodkumar, Eur. Phys. J. A \textbf{42}, 213 (2009).
		
		\bibitem{BernotasPRD} A. Bernotas, V. Simonis, Phys. Rev. D \textbf{87}, 074016 (2013).
		
		
		
		\bibitem{Albertus} C. Albertus, E. Hernandez, J. Nieves, J.M. Verde-Velasco, Eur. Phys. J. A \textbf{32}, 183 (2007).
		
		\bibitem{DhirPRD2021} A. Hazra, S. Rakshit and R. Dhir, Phys. Rev. D \textbf{104}, 053002 (2021).
		
		
		\bibitem{Li} H. S. Li, L. Meng, Z. W. Liu, S. L. Zhu, Phys. Lett. B
		\textbf{777}, 169–176 (2018).
		
		
	\end{thebibliography}
	

\end{document}